\documentclass[sigplan,screen,nonacm]{acmart}

\AtBeginDocument{%
  }

\usepackage{tikz}
\usepackage{listings}
\usepackage{balance}
\usetikzlibrary{trees}

\lstdefinestyle{terminalstyle}{
    backgroundcolor=\color{black!5},   
    commentstyle=\color{gray},
    keywordstyle=\color{blue},
    numberstyle=\tiny\color{gray},
    stringstyle=\color{teal},
    basicstyle=\footnotesize\ttfamily, 
    breakatwhitespace=false,
    breaklines=true,
    keepspaces=true,
    numbersep=5pt,
    showspaces=false,
    showstringspaces=false,
    showtabs=false,
    tabsize=2
}

\copyrightyear{2026}
\acmYear{2026}
\setcopyright{cc}
\setcctype{by}
\acmConference[ASPLOS '26] {Proceedings of the 30th ACM International Conference on Architectural Support for Programming Languages and Operating Systems, Volume 2}{March 21--26, 2026}{Pittsburgh, PA, USA.}
\acmBooktitle{Proceedings of the 30th ACM International Conference on Architectural Support for Programming Languages and Operating Systems, Volume 2 (ASPLOS '26), March 21--26, 2026, Pittsburgh, PA, USA}
\acmISBN{979-8-4007-2359-9/2026/03}
\acmDOI{10.1145/3779212.3790205}

\newcommand{\circlenumber}[1]{\raisebox{.5pt}{\textcircled{\raisebox{-.9pt} {#1}}}}
\newcommand{\subheading}[1]{\noindent \textbf{#1:}}
\newcommand{\revision}[1]{{#1}}

\author{Joshua Viszlai}
\affiliation{%
  \institution{University of Chicago}
  \department{Department of Computer Science}
  \city{Chicago}
  \state{IL}
  \country{USA}
}
\email{viszlai@uchicago.edu}

\author{Satvik Maurya}
\affiliation{%
  \institution{University of Wisconsin-Madison}
  \department{Department of Computer Sciences}
  \city{Madison}
  \state{WI}
  \country{USA}
}
\email{smaurya@wisc.edu}

\author{Swamit Tannu}
\affiliation{%
  \institution{University of Wisconsin-Madison}
  \department{Department of Computer Sciences}
  \city{Madison}
  \state{WI}
  \country{USA}
}
\email{swamit@cs.wisc.edu}

\author{Margaret Martonosi}
\affiliation{%
  \institution{Princeton University}
  \department{Department of Computer Science}
  \city{Princeton}
  \state{NJ}
  \country{USA}
}
\email{mrm@cs.princeton.edu}

\author{Frederic T. Chong}
\affiliation{%
  \institution{University of Chicago}
  \department{Department of Computer Science}
  \city{Chicago}
  \state{IL}
  \country{USA}
}
\email{chong@cs.uchicago.edu}

\begin{document}
\title{PropHunt: Automated Optimization of Quantum Syndrome Measurement Circuits}

\begin{abstract}
Fault-Tolerant Quantum Computing (FTQC) relies on Quantum Error Correction (QEC) codes to reach error rates necessary for large scale quantum applications. At a physical level, QEC codes perform parity checks on data qubits, producing syndrome information, through Syndrome Measurement (SM) circuits. 
These circuits define a code's logical error rate and must be run repeatedly throughout the entire program. The performance of SM circuits is therefore critical to the success of a FTQC system.

While ultimately implemented as physical circuits, SM circuits have challenges that are not addressed by existing circuit optimization tools. Importantly, inside SM circuits themselves errors are expected to occur, and how errors propagate through SM circuits directly impacts which errors are detectable and correctable, defining the code's logical error rate. This is not modeled in NISQ-era tools, which instead optimize for targets such as gate depth or gate count to mitigate the chance that any error occurs. This gap leaves key questions unanswered about the expected real-world effectiveness of QEC codes.

In this work we address this gap and present PropHunt, an automated tool for optimizing SM circuits \revision{for CSS codes}. We evaluate PropHunt on a suite of relevant QEC codes and demonstrate PropHunt's ability to iteratively improve performance and recover existing hand-designed circuits automatically. We also propose a near-term QEC application, Hook-ZNE, which leverages PropHunt's fine-grained control over logical error rate to improve Zero-Noise Extrapolation (ZNE), a promising error mitigation strategy. 

%
\end{abstract}

\ccsdesc[500]{Computer systems organization~Quantum computing}
\ccsdesc[500]{Hardware~Quantum error correction and fault tolerance}

\keywords{Quantum Computing; Quantum Error Correction; Syndrome Measurement Circuit; Error Mitigation}

\maketitle 

\section{Introduction}

\begin{figure}
    \centering
    \includegraphics[width=0.9\linewidth]{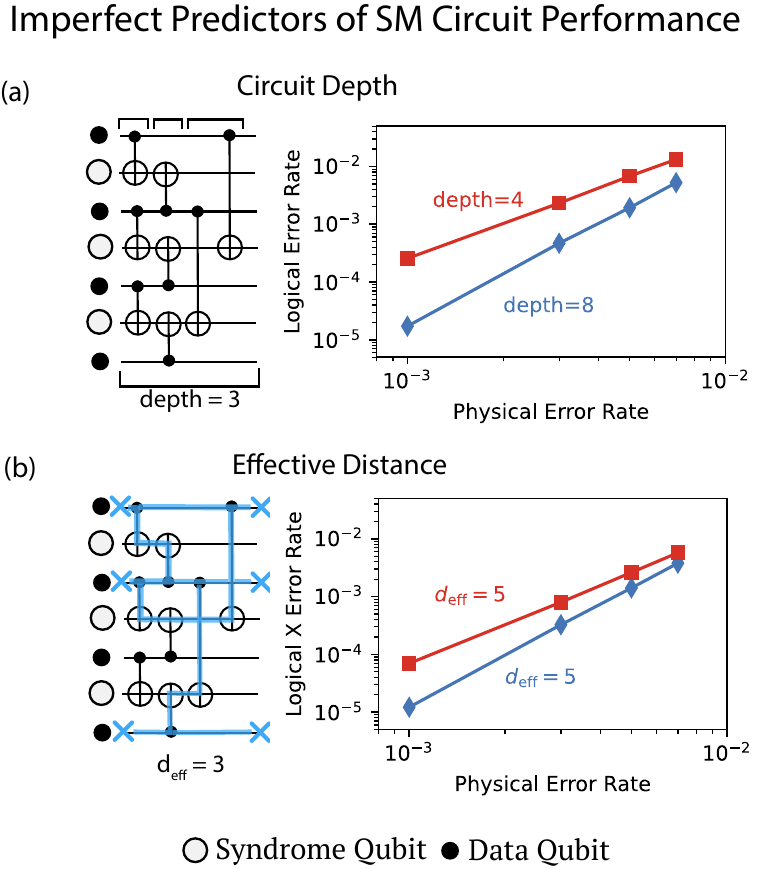}
    \caption{Common performance predictors targeted in Syndrome Measurement (SM) circuit design are imperfect. Data is from different SM circuits for a $d=5$ surface code. Red (square) plots indicate lower performing SM circuits despite having equal or better performance predictor values. Blue (diamond) plots indicate higher performing SM circuits despite having equal or worse performance predictor values.
    }
    \label{fig:counter_examples}
\end{figure}



Quantum Computing presents transformative potential to key applications in chemistry, material science, and cryptography. However, these applications require program sizes beyond the reach of noisy, physical quantum circuits.
To address this, researchers are focused on delivering Fault-Tolerant Quantum Computing (FTQC), where programs rely on an abstraction of Quantum Error Correction (QEC) to repeatedly detect and correct errors during execution.

QEC codes are the building blocks of FTQC, and recent years have seen an explosion in QEC code designs~\cite{hastings2021fiber, leverrier2022quantum, bravyi2024high, malcolm2025computing, lin2024quantum} and proposed implementations~\cite{Tremblay2022, bravyi2024high, xu2024constant, viszlai2023matching, poole2024architecture, pecorari2025high}.
While theoretical evaluations can inform promising QEC codes, full evaluations require a compilation to Syndrome Measurement (SM) circuits which implement a code's parity checks as a physical circuit, enabling the detection of errors.
Despite being key to QEC performance, techniques and tools for creating SM circuits in general are quite limited.
Part of this stems from the design space complexity of SM circuits. 
For moderate code sizes, these involve hundreds of qubits and gates. 
For example, the well known surface code with a code distance of $d=11$ can be implemented with a SM circuit using 241 qubits and 440 CNOT gates.
Additionally, SM circuits have specific constraints on CNOT ordering to preserve stabilizer commutation of the target QEC code.
This is not modeled in general-purpose, physical circuit optimization tools~\cite{zulehner2018efficient, li2019tackling, molavi2022qubit}.
Both of these features place SM circuit design beyond the reach of many existing approaches to circuit optimization.

Prior work has tamed the complexity of SM circuits in one of three ways. ~\circlenumber{1} hand-designed circuits tailored to specific codes~\cite{tomita2014low, Tremblay2022, lin2025single}. ~\circlenumber{2} brute-force searches over a reduced design space~\cite{beverland2021cost, bravyi2024high, lee2025color}. And ~\circlenumber{3} circuit synthesis targeting imperfect performance predictors~\cite{Shutty2022, Yin2025}.
However, none of these approaches are sufficient to address the growing space of QEC codes.
Relying on hand-designed circuits is intensive, and will not always be tractable.
For example, codes with random structure present little opportunity for hand-design.
Relying on brute-force searches is also unscalable.
While feasible for small codes, it quickly becomes intractable for the moderate or large code sizes necessary for FTQC.
Finally, we present counter examples, shown in Figure~\ref{fig:counter_examples}, to common performance predictors used in designing SM circuits: circuit depth and effective code distance. 
While circuit depth is important for coherence-limited systems, Figure~\ref{fig:counter_examples}(a) shows depth alone is insufficient to capture the resulting logical error rate. 
A minimum depth circuit that has poorly ordered CNOT gates creates problematic error propagation through a SM circuit.
This can then be outperformed by a circuit with well ordered CNOT gates, even if it is not minimum depth.
Figure~\ref{fig:counter_examples}(b) shows just using the effective code distance: the minimum number of errors needed to create an undetected logical error, is also insufficient. 
Performance gaps can exist between SM circuits with the same effective code distance.

Addressing these shortcomings can significantly improve resulting SM circuit quality.
For example, existing tools that target circuit depth without considering how errors propagate through CNOTs~\cite{Yin2025} can suffer a 10x increase in logical error rate for surface codes based on the data in Figure~\ref{fig:counter_examples}.

Our approach to SM circuit optimization has key differences compared to prior work.
\circlenumber{1} We optimize SM circuits at the granularity of individual CNOTs, exploring a fuller design space.
\circlenumber{2} We efficiently search through the larger design space by identifying and minimizing \textit{ambiguity} in the SM circuit.
Based on algorithmic insights, we introduce a generally applicable tool, PropHunt, for automated optimization of SM circuits.
Applying our tool to surface codes, we show PropHunt is able to automatically recover the performance of hand-designed circuits.
Furthermore, we find applying PropHunt to specific Lifted Product (LP) codes and Random Quantum Tanner (RQT) codes successfully optimizes SM circuit performance, reducing logical error rates by 2.5x-4x compared to a standard coloration circuit~\cite{Tremblay2022} at a physical error rate of 0.1\%.
We additionally motivate a near-term QEC application of PropHunt.
We show intermediate SM circuits created by PropHunt can serve as useful, low-overhead error mitigation gadgets with Zero-Noise Extrapolation (ZNE) which we term Hook-ZNE.
Our evaluations show Hook-ZNE improves over existing QEC+ZNE solutions, reducing error by 3x-6x compared to Distance-Scaling ZNE~\cite{wahl2023zero}.

We summarize our contributions as follows:
\begin{itemize}
    \item A novel solution for finding min-weight logical errors in a circuit-level decoding graph, allowing for efficient, parallelized circuit optimization. 
    \item An approach for optimizing SM circuits based on modifying error propagation to minimize ambiguity, directly addressing logical error rate rather than using imperfect performance predictors.
    \item PropHunt, a tool for optimizing SM circuits based on our approach that minimizes ambiguity.
    \item SM circuits for select Lifted Product and Random Quantum Tanner codes that have 2.5x-4x lower logical error rates compared to known SM circuits.
    \item An additional application of PropHunt, Hook-ZNE, that uses intermediate SM circuits to improve Zero-Noise Extrapolation for near-term FTQC circuits.
\end{itemize}


\section{Background}

For a broader introduction to quantum computing, see~\cite{ding2020quantum, nielsen2001quantum, rieffel2011quantum}.  
Here, we focus on the essentials of quantum error correction (QEC) relevant to this work—stabilizer codes, their measurement circuits, and the role of gate ordering in preserving error resilience.

\subsection{Quantum Stabilizer Codes}

An $[[n,k,d]]$ QEC code encodes $k$ logical qubits into $n$ physical qubits.  
The \emph{code distance} $d$ is the minimum number of physical qubit errors that can cause an undetectable logical error, equivalently the weight of the smallest logical operator.

The surface code is a widely studied stabilizer code defined on a two-dimensional qubit lattice.  
Data qubits store the encoded quantum information, while \emph{syndrome qubits} measure multi-qubit parity checks, known as \emph{stabilizers}, which detect the presence of errors without collapsing the logical state.  
A rotated surface code of distance $d$ contains $d^2$ data qubits and $d^2-1$ syndrome qubits, evenly divided between $X$-type stabilizers (which detect $Z$ errors) and $Z$-type stabilizers (which detect $X$ errors).

\subsection{Example: $d=3$ Rotated Surface Code}
\begin{figure}[h]
    \centering
    \includegraphics[width=0.6\linewidth]{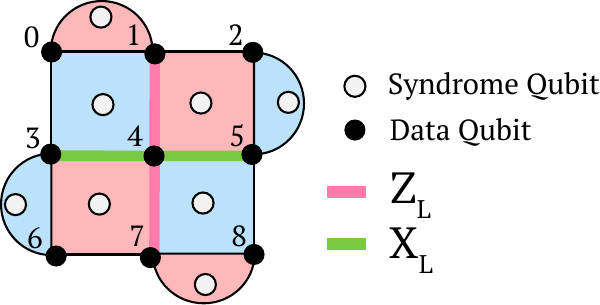}
    \caption{A $d=3$ surface code. Blue faces are $X$-type stabilizers and red faces are $Z$-type stabilizers. }
    \label{fig:d3_surface}
\end{figure}
Consider the $d=3$ rotated surface code, shown in Figure~\ref{fig:d3_surface}.
The code has nine data qubits arranged in a $3\times 3$ grid and eight syndrome qubits implementing four $X$-type and four $Z$-type stabilizers.  
These can be expressed as parity-check matrices $H_X$ and $H_Z$:

\setlength{\arraycolsep}{2pt} 
\begin{small}
\[
H_X =
\left[
\begin{array}{*{9}{c@{\hskip 3pt}}}
1&1&0&1&1&0&0&0&0\\
0&0&0&0&1&1&0&1&1\\
0&0&0&1&0&0&1&0&0\\
0&0&1&0&0&1&0&0&0
\end{array}
\right],
\quad
H_Z =
\left[
\begin{array}{*{9}{c@{\hskip 3pt}}}
0&1&1&0&1&1&0&0&0\\
0&0&0&1&1&0&1&1&0\\
1&1&0&0&0&0&0&0&0\\
0&0&0&0&0&0&0&1&1
\end{array}
\right]
\]
\end{small}
Each row corresponds to a stabilizer, and a $1$ in column $j$ indicates that the stabilizer acts on data qubit $j$.

\subsection{From Check Matrices to Ideal SM Circuits}

\begin{figure}[h]
    \centering
    \includegraphics[width=\linewidth]{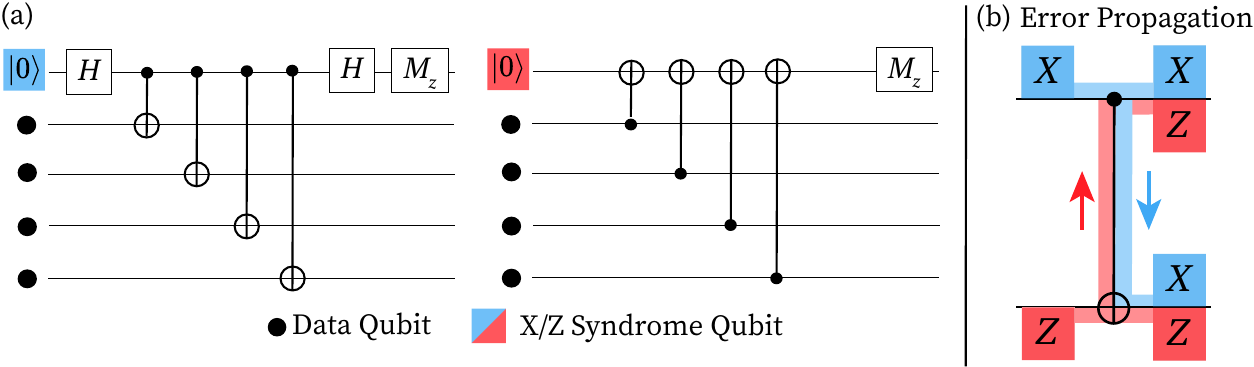}
    \caption{(a) Example syndrome measurement circuits for measuring an X check (left) and a Z check (right). (b) CNOT error propagation rules.}
    \label{fig:stab_circ}
\end{figure}
The matrices $H_X$ and $H_Z$ directly specify a syndrome measurement (SM) circuit.  
As shown in Figure~\ref{fig:stab_circ}a, for an $X$-type stabilizer (row of $H_X$), the corresponding circuit prepares an ancilla in the $|+\rangle$ state, applies a CNOT from the ancilla (control) to each involved data qubit (target), and measures the ancilla in the $X$ basis.  
For a $Z$-type stabilizer (row of $H_Z$), the ancilla is prepared in $|0\rangle$, and for each involved data qubit, a CNOT is applied with the data qubit as control and the ancilla as target, followed by a $Z$-basis measurement of the ancilla.  

Running all stabilizer circuits produces the syndrome vectors
\[
s_X = H_X e_Z \ (\mathrm{mod}\ 2), \quad s_Z = H_Z e_X \ (\mathrm{mod}\ 2),
\]
where $e_Z$ and $e_X$ are binary vectors indicating $Z$ and $X$ errors on the data qubits, respectively.

\subsection{From Ideal SM Circuits to Logical Errors}

A \textit{decoder} uses the syndrome vectors $s_X$ and $s_Z$ to try and deduce the errors $e_Z$ and $e_X$. 
It can then infer the state of the logical qubits through the logical observable matrices $L_X$ and $L_Z$. 
For the $d=3$ rotated surface code these are:

\setlength{\arraycolsep}{2pt} 
\begin{small}
\[
L_X =
\left[
\begin{array}{*{9}{c@{\hskip 3pt}}}
0&0&0&1&1&1&0&0&0
\end{array}
\right],
\quad
L_Z =
\left[
\begin{array}{*{9}{c@{\hskip 3pt}}}
0&1&0&0&1&0&0&1&0
\end{array}
\right]
\]
\end{small}
Then computing
\[
l_Z = L_X e_Z \ (\mathrm{mod}\ 2), \quad l_X = L_Z e_X \ (\mathrm{mod}\ 2),
\]
results in the logical error vectors $l_Z$ and $l_X$, indicating if a logically encoded qubit experienced a Z or X error, respectively.

\subsection{Syndrome Generation and Decoding Examples}

We illustrate the process of syndrome generation and decoding for $X$ errors, which are detected by $H_Z$.  

\paragraph{Correctable Error.}
A single $X$ error on data qubit $5$ corresponds to
$
e_X = (0,0,0,0,1,0,0,0,0)^T.
$
Since qubit $5$ participates in rows $1$ and $2$ of $H_Z$, the resulting syndrome is
$
s_Z = (1,1,0,0)^T,
$
which uniquely identifies qubit $5$ as faulty. 

The decoder determines the logical error is
$
l_X = (1)^T,
$
indicating a logical X error occurred.
The decoder can then flip the logical qubit back, fully restoring the code state.

\paragraph{Uncorrectable Error.}
Consider $X$ errors on qubits $\{1,5,9\}$:
$
e_X = (1,0,0,0,1,0,0,0,1)^T.
$
Here, 
$
s_Z = (0,0,0,0)^T,
$
indicating the pattern is undetected by all stabilizers.
However
$
l_X = (1)^T,
$
indicating the pattern results in a logical X error. 
Since an undetected logical error was created, the code state cannot be restored.

\subsection{Noisy SM Circuits}

In practice, the CNOT gates in SM circuits are imperfect: faults may occur on either the data or ancilla qubits at any point during the circuit.  
The fundamental property of QEC is that as long as the number and arrangement of such faults remain below the code's threshold, the decoder can still recover the logical state.

Error propagation through a CNOT  (Figure~\ref{fig:stab_circ}b) follows deterministic rules:
\begin{align*}
X_c &\rightarrow X_c X_t, & Z_c &\rightarrow Z_c,\\
X_t &\rightarrow X_t,     & Z_t &\rightarrow Z_c Z_t,
\end{align*}
where $c$ denotes the control and $t$ the target qubit.  
These rules explain how data errors can flip stabilizer outcomes (desirable for detection), but also how ancilla errors, called hook errors, can spread to multiple data qubits (potentially harmful).

\subsection{The Circuit-Level Model}\label{sec:circuit_level}

To capture the behavior of noisy SM circuits, a \textit{circuit-level} model is used.
As gate errors can often be modeled as random Pauli gates, a circuit-level model statically \textit{propagates} errors through a SM circuit to deduce the set of syndromes each error will flip.
As a result, in the circuit-level model the code-level check and logical observable matrices are replaced by circuit-level counterparts, $H$ and $L$, that replace qubits with gate errors. 
$H$ maps gate errors (columns) to the set of syndromes they flip (rows), and $L$ maps the same gate errors (columns) to the set of logical observables they flip (rows).

\begin{figure}
    \centering
    \includegraphics[width=0.8\linewidth]{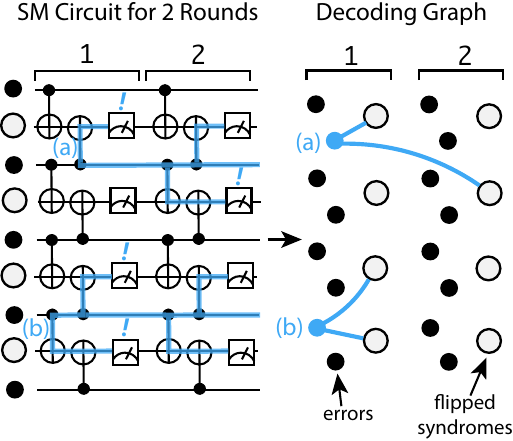}
    \caption{A circuit-level model for a noisy SM circuit. Decoding graph edges are only drawn for example errors (a) and (b).}
    \label{fig:circuit-level}
\end{figure}

As shown in Figure~\ref{fig:circuit-level}, the ordering of CNOTs influences when errors are first detected.
For example, an $XX$ CNOT error, (a), is detected by one syndrome qubit in the first round and a different syndrome qubit in the second round. 
Due to different error propagation, another $XX$ CNOT error, (b), is detected by two syndrome qubits in the first round.

Importantly, since different SM circuits propagate errors differently, they correspond to different circuit-level check matrices, $H$ and $L$.

Circuit-level matrices often include errors from $d$ rounds of an SM circuit, and as a result can be significantly larger than the original stabilizer and logical observable matrices.
For example, stabilizer matrices $H_Z$ and $H_X$ each have $49$ columns for a $d=7$ surface code, but the circuit-level check matrix $H$ can have $>15,000$ columns corresponding to possible errors.

\subsection{Hook Errors}

A \emph{hook error} occurs when an ancilla fault in an SM circuit propagates to multiple data qubits via the CNOT error propagation rules.  
Consider a $Z$-type check on qubits $\{q_1,q_2,q_3,q_4\}$, with the ancilla as the target of all CNOTs.  
A $Z$ error on the ancilla after the second CNOT will propagate back to the control data qubits of subsequent CNOTs, creating a correlated multi-qubit fault.  
If these qubits overlap with a minimum-weight logical operator, the fault can reduce the \emph{effective code distance} and cause a logical error with fewer than $d$ faults.

\begin{figure}[h]
    \centering
    \includegraphics[width=\linewidth]{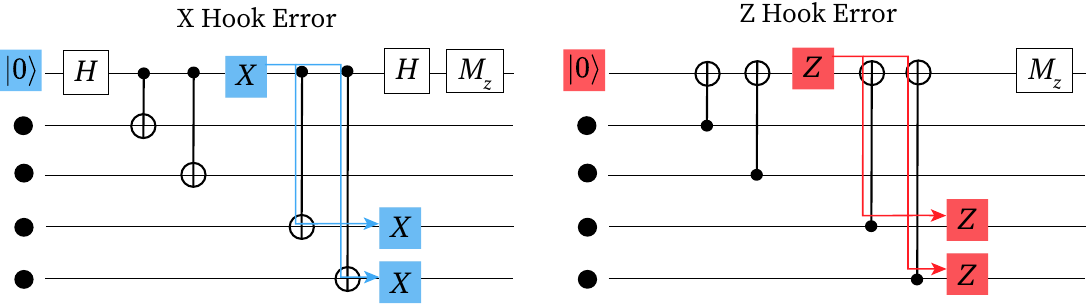}
    \caption{Worst case X and Z hook errors for weight 4 checks.}
    \label{fig:hook}
\end{figure}

We can view a hook error on a weight $w$ check as applying part of the corresponding stabilizer operator and so a hook error spread to $w' > \lfloor \frac{w}{2} \rfloor$ data qubits can instead be interpreted as an error on $w - w'$ data qubits by multiplying by the stabilizer operator. As a result in the worst case a hook error can cause a single error to become a multi-qubit error on $\lfloor \frac{w}{2} \rfloor$ data qubits, as shown in Figure~\ref{fig:hook}.

\subsection{Effective Code Distance}\label{sec:d_eff}

The effective code distance $d_{\text{eff}}$ is the minimum number of physical faults that create an undetected logical error for a given SM circuit.  
Poor CNOT ordering can reduce $d_{\text{eff}}$ by allowing early ancilla faults to overlap with logical operators.  
Carefully chosen gate schedules can prevent such overlaps, ensuring $d_{\text{eff}} = d$.

While Figure~\ref{fig:counter_examples} shows $d_{\text{eff}}$ is not a perfect predictor of performance by itself, it is still meaningful. 
An SM circuit with a lower $d_{\text{eff}}$ can be expected to perform worse, and determining $d_{\text{eff}}$ is useful in explaining SM circuit performance.

\section{Motivation}
\begin{figure}
    \centering
    \includegraphics[width=\linewidth]{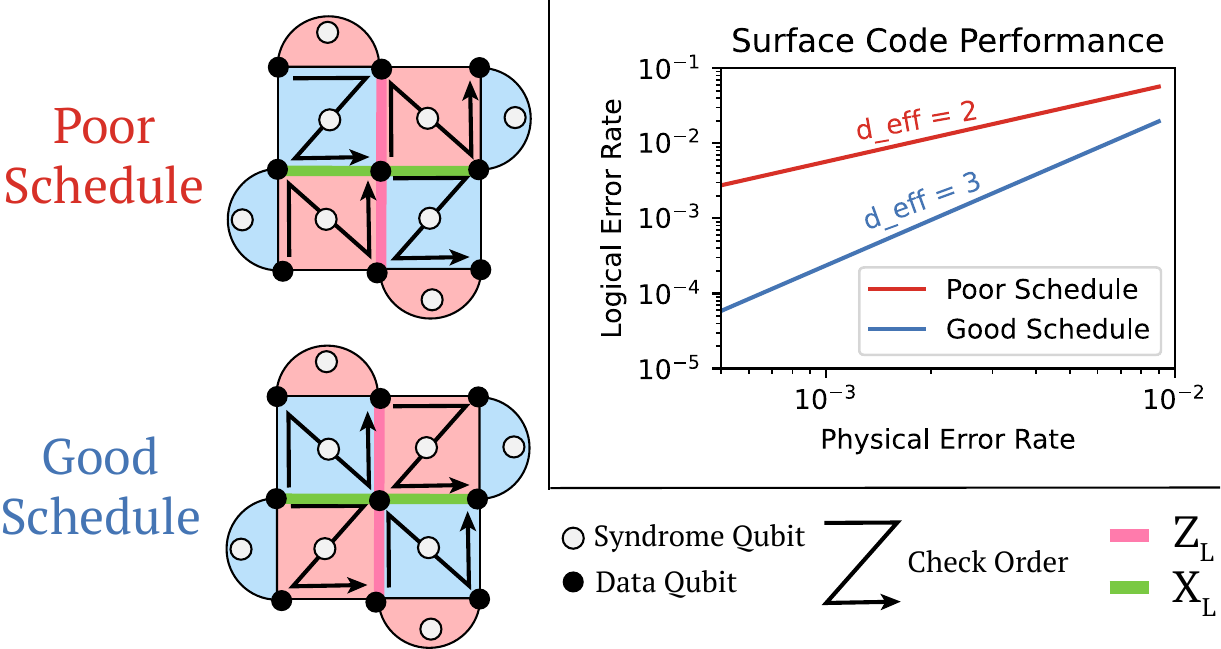}
    \caption{Comparing the logical performance of a $d=3$ surface code when using two different CNOT schedules. The check order indicates the order a syndrome qubit performs CNOTs with its connected data qubits.}
    \label{fig:surface_hook}
\end{figure}

Currently there is a lack of approaches to optimize SM circuits. 
This can leave designers in the dark about the actual effectiveness of their QEC code.
NISQ-era tools for optimizing circuits focus on minimizing gate counts and circuit depth to reduce the chance of physical errors occurring. 
For SM circuits, however, errors are \textit{expected to occur} and must be tolerated.
This presents a novel challenge as a SM circuit dictates how errors \textit{propagate} into syndrome patterns that are then decoded.
However, generally-applicable, automated solutions for SM circuit design are limited.
Instead research on a QEC code where a circuit-level analysis is present will often have a section dedicated to how SM circuits are designed. 
Such solutions are generally hand-tailored to the code of interest and not applicable to other codes. 
Furthermore, as we point out in Figure~\ref{fig:counter_examples}, common targets used in SM circuit design are imperfect predictors of circuit performance.
These features not only inhibit research progress, as designing SM circuits is a non-trivial task that must be redone for each code, but also inhibits code comparisons. 
Current attempts to compare codes at a circuit-level can struggle to disentangle whether performance differences are caused by a chosen SM circuit or by a code itself.

Despite the current lack of tooling, it's well known that SM circuits can have a significant influence on a code's logical error rate. 
Here, we present a motivating example of this and discuss how prior work has found performant SM circuits.
\subsection{Motivating Example: Surface Code}
The surface code has a well known schedule of CNOTs for a good SM circuit often referred to as the 'N-Z' schedule~\cite{tomita2014low}. 
This schedule relies on the observation that any logical X (Z) operator in the surface code is horizontal (vertical). 
The 'N-Z' schedule therefore orders the CNOTs between check qubits and data qubits such that the worst case X (Z) hook errors are perpendicular to the X (Z) logical operator.

Figure~\ref{fig:surface_hook} compares two SM circuits for a $d=3$ surface code: one that does not use the proper 'N-Z' schedule and one that does. 
The check order indicates the order a syndrome qubit performs CNOTs with its connected data qubits.
We can see that using the proper 'N-Z' schedule has a notable impact on the code's resulting logical error rate and that using a poor schedule can reduce $d_{\text{eff}}$.
However, the 'N-Z' schedule is a hand-designed schedule for a SM circuit.
In general, we would like an automated way to \circlenumber{1} recover the 'N-Z' schedule for the surface code, and \circlenumber{2} find similarly performant schedules for other QEC codes.

\revision{
Interestingly, the sensitivity of logical error rate to different SM circuits is code-specific. As an example, for hypergraph-product codes it's known that all SM circuits have $d_{\text{eff}}=d$~\cite{manes2025distance}. Alternatively for the Steane code, we can note that all CNOT orderings produce hook errors that are distance-reducing, necessitating alternative approaches to SM circuit design. For most QEC codes, however, variance in logical error rate for possible SM circuits is unknown, motivating an automated approach for finding SM circuits.
}





 
\begin{figure*}
    \centering
    \includegraphics[width=0.75\textwidth]{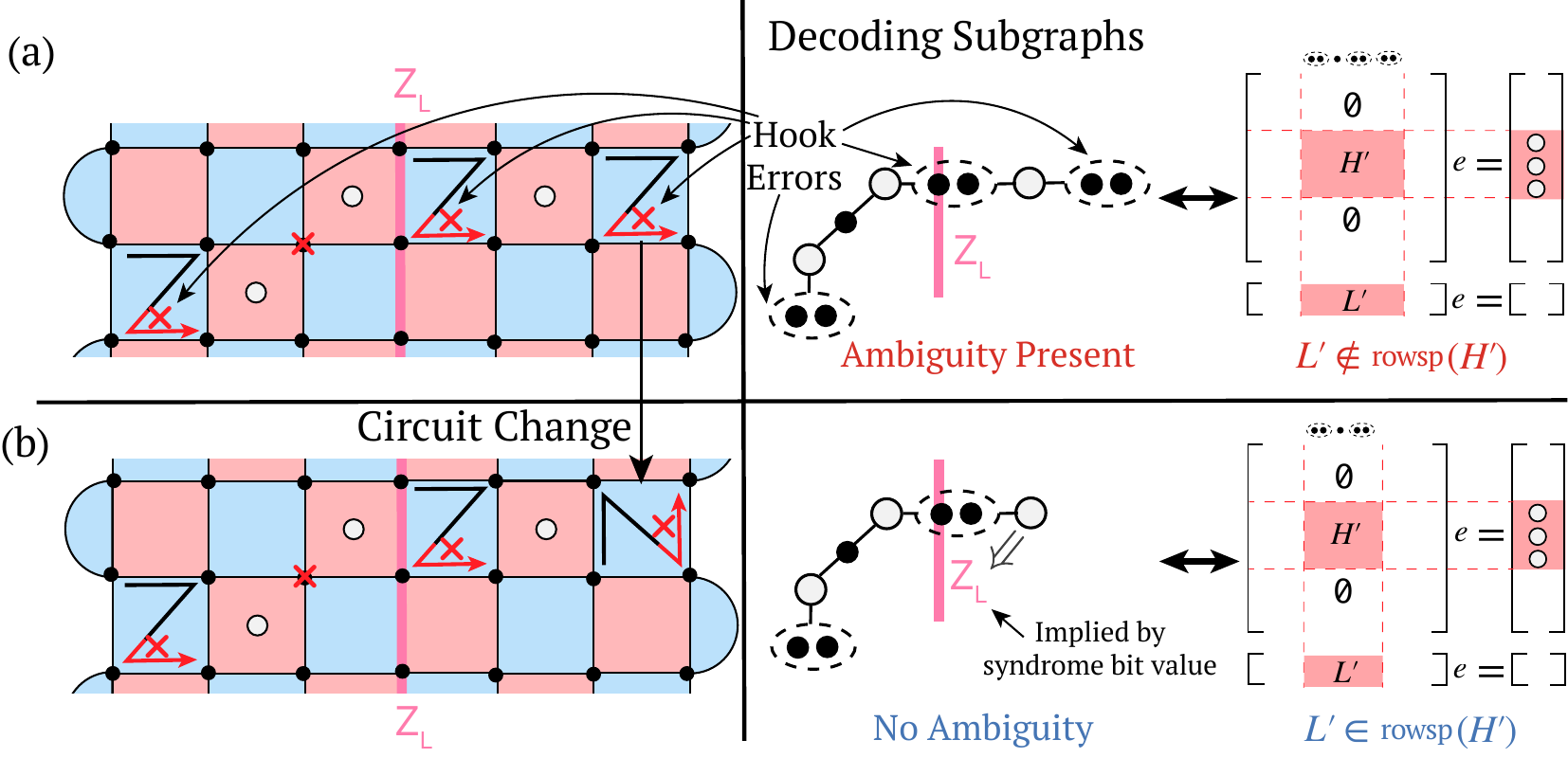}
    \caption{A $d=7$ surface code with one CNOT ordering (a) and a modified CNOT ordering (b) from the perspective of ambiguity. A logical error in (a) corresponds to a decoding subgraph containing ambiguity. Any assignment of syndrome bit values that is decodable on the subgraph can be decoded in two ways that each result in a different effect on $Z_L$. The same decoding subgraph in (b) no longer has ambiguity as the state of $Z_L$ is implied by the right-most syndrome bit. We can further verify a logical error has been removed by checking that the updated error mechanisms don't correspond to a new logical error on different syndrome bits.}
    \label{fig:surface_ambiguity}
\end{figure*}

\section{Error Ambiguity Explained}\label{sec:faults}

Our approach views logical errors as stemming from \textit{ambiguity}. 
Here we provide a formal definition of ambiguity and a worked out example for the surface code.

While the effective code distance $d_{\text{eff}}$ describes a minimum weight logical operator at the circuit-level, the logical error rate scales as $O(p^{\lceil d_{\text{eff}}/2 \rceil})$. 
Logical faults occur due to the decoder \textit{mispredicting} the state of the logical qubits as a result of \textit{ambiguous errors}. 
Formally, we can define ambiguous errors as two sets of errors, $e_1$ and $e_2$, such that $He_1 = He_2$ and $Le_1 \neq Le_2$. 
In other words, two error patterns that have the same syndromes but different logical effects. 
A most likely error (MLE) decoder will therefore predict the most likely of the two to have occurred.
As a result, mispredictions will occur with a probability proportional to the least likely of the two errors. 
We can further note that the union, $e_L = e_1 + e_2\ (\mathrm{mod}\ 2)$, is an undetected logical error since $He_{L} = He_1 + He_2 = 0\ (\mathrm{mod}\ 2)$, and $Le_L = Le_1 + Le_2 \neq 0\ (\mathrm{mod}\ 2)$. 
Therefore a logical error of weight $d_{\text{eff}}$ corresponds to many different sets of ambiguous errors, and closely balanced ambiguous errors, $e_1$ with weight $w_1 = \lfloor d_{\text{eff}} / 2 \rfloor$ and $e_2$ with weight $w_2 = \lceil d_{\text{eff}} / 2 \rceil$, are the most likely failure cases.

\subsection{Checking Ambiguity}\label{sec:ambiguity}
As discussed in~\cite{wolanski2024ambiguity}, the presence of ambiguous errors can be detected efficiently. 
Given a subset of syndromes $s'$ there can be an ambiguous explanation if $L' \notin \text{rowsp}(H')$, where $H'$ and $L'$ are sub-matrices of $H$ and $L$ corresponding to errors connected only to the syndromes $s'$. 

A key insight is since $H$ and $L$ change with the SM circuit, as discussed in Section~\ref{sec:circuit_level}, we can check if for some modified SM circuit $L' \in \text{rowsp}(H')$, allowing $s'$ to be decoded unambiguously and removing a logical fault pathway. 

\subsection{Comparing Ambiguity in SM Circuits}
Figure~\ref{fig:surface_ambiguity}a shows a logical error in a $d=7$ surface code corresponding to a reduced code distance of 4 due to hook errors from poorly scheduled CNOTs.
Looking at the subgraph of the circuit-level decoding graph corresponding to the errors involved, we can see the logical error exists due to the possibility of ambiguity. 
Any assignment of syndrome bit values that is decodable on the subgraph can be decoded in two ways that each result in a different effect on $Z_L$. 
For example, the middle syndrome being flipped can be explained by either the two hook errors on the right or the single error and hook error combination on the left. 
As both are equally likely, the decoder will not be able to reliably determine which case occurred. 
And since the first case flips the Z observable and the second does not, this will lead to a logical error.

In Figure~\ref{fig:surface_ambiguity}b we can see that for a modified CNOT ordering, the same decoding subgraph no longer has the possibility of ambiguous errors. 
The effect on $Z_L$ is implied by the right-most syndrome bit for any assignment of syndrome bit values decodable on the subgraph.
We can further verify we've removed a potential logical error by checking that the updated errors, the three hook errors and single error, are also no longer a logical error.


\begin{figure*}
    \centering
    \includegraphics[width=\textwidth]{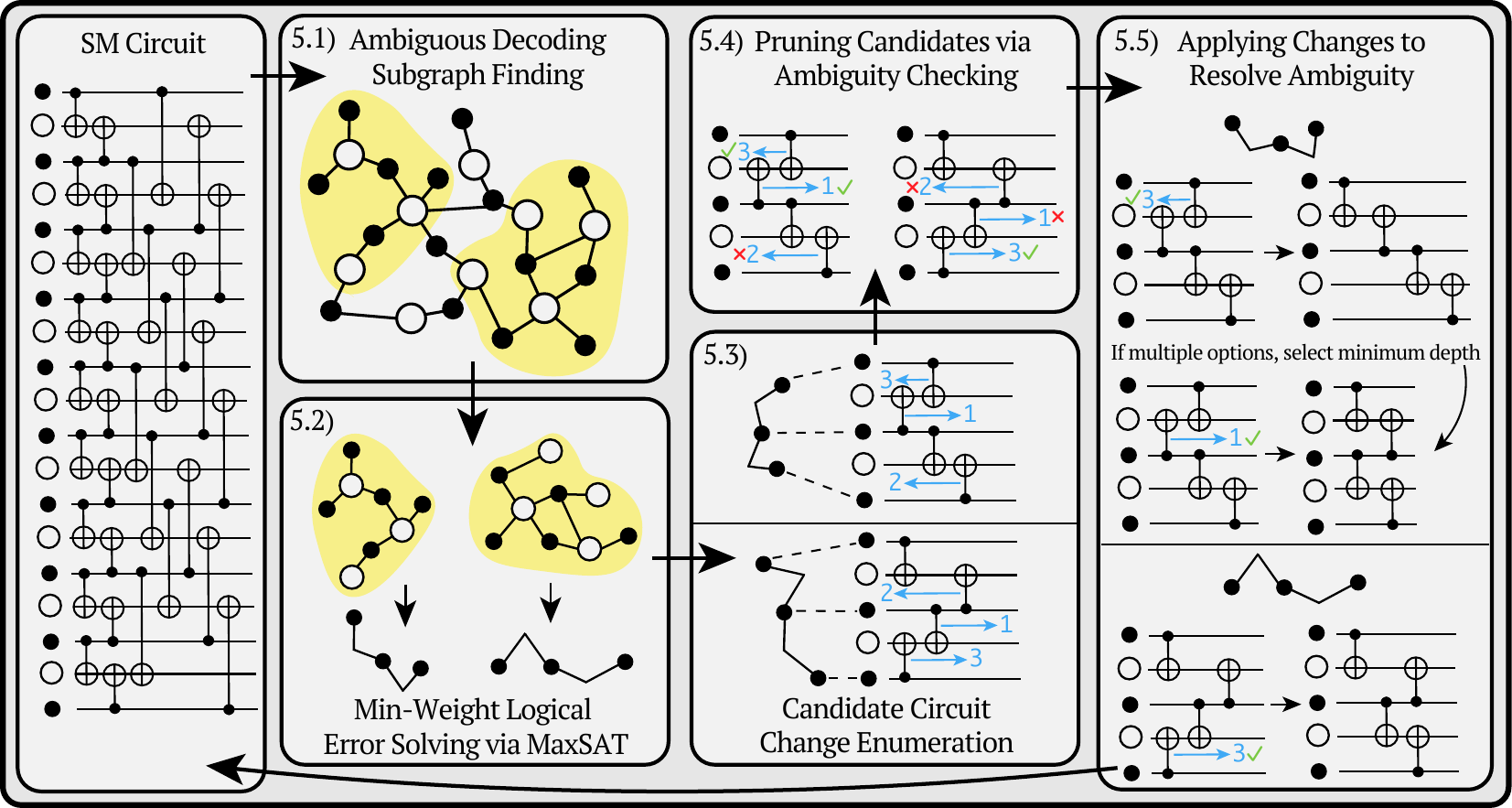}
    \caption{An overview of PropHunt's iterative optimization loop.}
    \label{fig:overview}
\end{figure*}

\section{PropHunt}

In this section we introduce a novel approach for automated optimization of SM circuits. 
Figure~\ref{fig:overview} gives an overview of our approach. 
First, a circuit-level decoding graph is produced from the SM circuit. 
Second, random subgraphs are expanded until $L' \notin \text{rowsp}(H')$ indicating the presence of ambiguous errors. 
Third, min-weight logical errors are found in each subgraph using a MaxSAT formulation. 
Fourth, SM subcircuits corresponding to each logical error are modified to change the error propagation and remove the ambiguity. 

\revision{
An end-to-end example of this idea can be understood using the surface code. Starting from the poor schedule shown in Figure~\ref{fig:d3_surface}, our goal is to iteratively identify subgraphs shown in Figure~\ref{fig:surface_ambiguity} and apply circuit changes that resolve the ambiguity, such as the one in Figure~\ref{fig:surface_ambiguity}b. Done repeatedly, a SM circuit of comparable performance to the good schedule shown in Figure~\ref{fig:d3_surface} should be produced.
}


\subsection{Ambiguous Decoding Subgraph Finding}\label{sec:ambig_finding}
We operate on bipartite circuit-level decoding graphs with sets of nodes corresponding to syndromes and errors. 
An edge between a syndrome node and an error node indicates that error flips that syndrome. 
To find possible ambiguous errors, we begin by selecting random error nodes as starting points. 
We additionally add all syndrome nodes connected to the error node to the subgraph. 
At each step we expand the subgraph by adding an additional error node and its connected syndrome nodes.
Furthermore, any error nodes connected only to syndrome nodes already in the subgraph are automatically included.
Subgraphs correspond to submatrices $H'$ and $L'$ and so after each step we check if $L' \notin \text{rowsp}(H')$, indicating whether the subgraph contains an ambiguous error. 
If an ambiguous error is present in a subgraph we halt its expansion.

We note that subgraphs containing disconnected components cannot contain a single ambiguous error spanning more than one connected component. 
During expansion, we therefore restrict ourselves to connected subgraphs--only error nodes adjacent to an already connected syndrome node are considered. 


\subsection{Solving for Minimum Weight Errors}\label{sec:max_sat}
MaxSAT is a useful framework in solving for minimum weight logical errors in QEC. 
Here we describe how PropHunt uses MaxSAT to solve for min-weight logical errors in ambiguous subgraphs. 
Our formulation is similar to the approach in Stim~\cite{gidney2021stim} for finding minimum weight logical errors.

For a subgraph with error nodes $E'$, syndrome nodes $S'$, and submatrices $H'$, $L'$, we first define variables $E'_i$ and $S'_i$ for all nodes in the subgraph. 
We then define
\begin{align*}
    S'_{i} = E'_j\otimes ... \otimes E'_k \text{ where } H_{i,j}=1, .., H_{i,k}= 1 
\end{align*}
as a set of hard constraints enforcing syndromes as the parity of their connected errors.
Similarly, we can define logical observable variables $L'_i$ and a set of hard constraints
\begin{align*}
    L'_{i} = E'_j\otimes ... \otimes E'_k \text{ where } L_{i,j}=1, .., L_{i,k}= 1 
\end{align*}
enforcing the logical observables as the parity of their connected errors.

Naively converting multivariate XOR clauses into Conjunctive Normal Form (CNF) is known to hinder solver performance as the resulting clause is exponential in size~\cite{prestwich2021cnf}.
To avoid this, we introduce auxiliary variables ($a_i$) in a standard, tree-like structure: 
\begin{center}
\begin{tikzpicture}[level distance=1cm,
  level 1/.style={sibling distance=3cm},
  level 2/.style={sibling distance=1.5cm}]
  \node {$S_i' = a_1 \otimes a_2$}
    child {node {$a_1 = E'_1 \otimes E'_2$ }
      child {node {$E'_1$}}
      child {node {$E'_2$}}
    }
    child {node {$a_2 = E'_3 \otimes E'_4$}
    child {node {$E'_3$}}
      child {node {$E'_4$}}
    };
\end{tikzpicture}
\end{center}

We then add additional hard constraints enforcing that the assigned error pattern is undetected by all stabilizers
\begin{align*}
    \bigvee_{i} S'_i = False
\end{align*}
and flips at least one logical observable
\begin{align*}
    \bigvee_{i} L'_i = True
\end{align*}

Finally we add soft constraints for each error $E_i$
\begin{align*}
    E_i = False
\end{align*}
so the solver optimizes for the fewest error assignments while satisfying the hard constraints.
This produces an optimal solution corresponding to a minimum weight logical error in the subgraph.

To solve, we use Z3~\cite{de2008z3} to simplify the model and to apply the Tseitin transformation~\cite{tseitin1983complexity} to convert to CNF. 
We then use the Loandra MaxSAT solver~\cite{berg2019core} with a timeout of 360 seconds \revision{run on a Intel Xeon Silver 4116 CPU core.}

\subsection{Candidate SM Circuit Change Enumeration}\label{sec:change}

As described in Section~\ref{sec:circuit_level}, in a circuit-level noise model errors occur during gates in the SM circuit. 
We can therefore map an error back to the gate that caused it. 
Using this, for each found minimum-weight logical error we enumerate a set of candidate SM circuit changes based on the involved gate errors. 
We implement two types of circuit changes, both of which modify the propagation order and the corresponding set of syndromes flipped by the circuit-level error.

\subsubsection{Reordering Changes}\label{sec:reorder}

We define \textit{reordering} as changing the order of data qubits a syndrome qubit interacts with. 
This changes the set of data qubits involved in hook errors which then corresponds to a different set of flipped syndrome bits.
\begin{figure}[h]
    \centering
    \includegraphics[width=\linewidth]{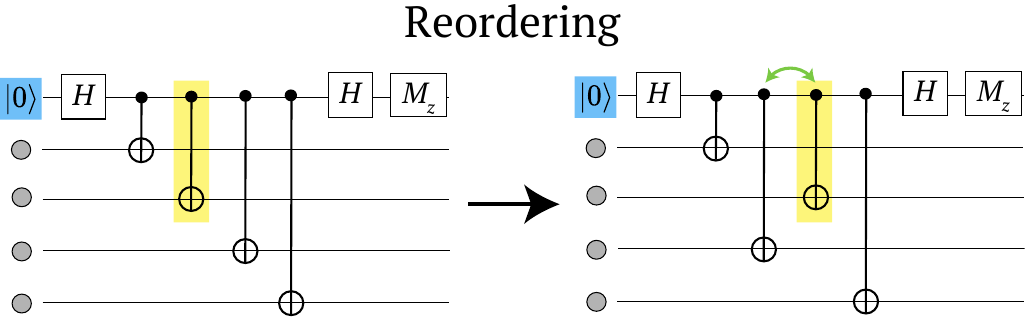}
    \caption{A reordering change to modify the behavior of a circuit-level error corresponding to the highlighted CNOT.}
    \label{fig:reordering}
\end{figure}

If the circuit-level error being modified is a hook error we enumerate $w - 1$ reordering changes
\begin{align*}
\{q_1 \ldots \boldsymbol{q_i} \ldots q_w\} \rightarrow \{q_1 \ldots \boldsymbol{q_j, q_i} \ldots q_w\} \forall j \in \{1\ldots w\} \setminus \{i\}
\end{align*}
where $w$ is the stabilizer weight and $q_i$ is the data qubit involved in the CNOT that caused the hook error. 
Each of these changes corresponds to reordering another data qubit, $q_j$, before $q_i$. 

\subsubsection{Rescheduling Changes}
We define \textit{rescheduling} as changing the relative scheduling of syndrome qubits on a shared data qubit. 
This changes the timesteps at which circuit-level errors are detected on syndrome qubits.
\begin{figure}[h]
    \centering
    \includegraphics[width=\linewidth]{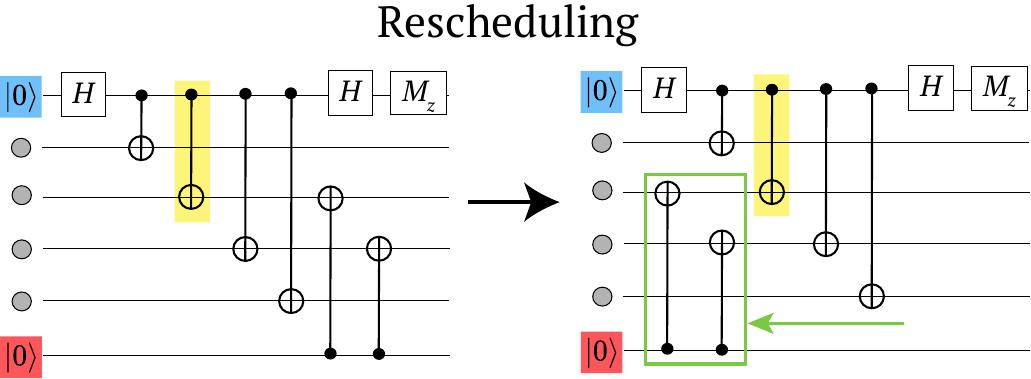}
    \caption{A rescheduling change to modify the behavior of a circuit-level error corresponding to the highlighted CNOT. Since opposite type (X/Z) stabilizers are involved, a second CNOT needs to be rescheduled to preserve commutation.}
    \label{fig:rescheduling}
\end{figure}

For a circuit-level error originating from a CNOT between data qubit $q_i$ and syndrome qubit $s_j$, we enumerate $|S_{q,i}|$ rescheduling changes, where $|S_{q,i}|$ is the set of syndrome qubits flipped by the circuit-level error on $q_i$. 
For each syndrome qubit $s_i \in S_{q,i}$, we swap the relative ordering of $s_i$ and $s_j$ on qubit $q_i$.
If $s_i$ and $s_j$ correspond to a pair of X and Z stabilizers, than we need to make an additional rescheduling change to preserve stabilizer commutation.
We select another qubit $q_k$ connected to both $s_i$ and $s_j$ and additionally swap the relative ordering of $s_i$ and $s_j$ on it.
For many codes, such as the surface code, there are uniquely two qubits $q_i, q_k$ shared between X and Z stabilizers, and so the selection of $q_k$ is deterministic.
Otherwise, $q_k$ is selected at random.

\begin{figure}
    \centering
    \includegraphics[width=\linewidth]{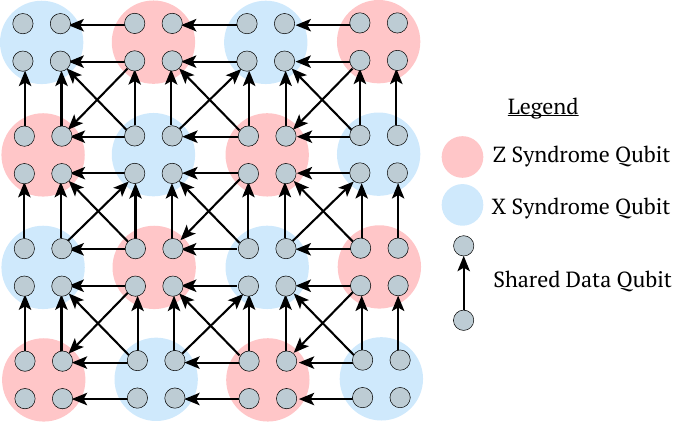}
    \caption{Internal representation used in PropHunt to track relative scheduling of syndrome qubits on shared data qubits.}
    \label{fig:prop_graph}
\end{figure}

Internally, PropHunt uses a multi-edge, directed graph representation (Figure~\ref{fig:prop_graph}) to simplify the tracking of rescheduling changes. 
Each syndrome qubit corresponds to a node in the graph, and an edge is added between pairs of nodes for each shared data qubit.
The direction of edges indicate which syndrome qubit interacts with the data qubit first.
This implicitly models the propagation of errors, as an out edge from a syndrome node indicates a CNOT error between the syndrome and data qubit pair can propagate to the connected syndrome node \textit{the same round}.
In edges similarly indicate a CNOT error can propagate to the connected syndrome node \textit{the next round}.

With this representation, rescheduling changes amount to flipping the direction of edges.
This graph coupled with specified syndrome qubit orderings as described in Section~\ref{sec:reorder}, define CNOT ordering dependencies that can then be directly converted to a CNOT schedule.

\subsection{Pruning SM Circuit Changes}\label{sec:prune}
In the previous stage we produce $O(wd_{\text{eff}})$ candidate changes to the SM circuit for each minimum weight-logical error found.
Most of these changes will not improve the code's logical error rate.
To prune the set of candidate changes, we implement two checks:

\textbf{Circuit Validity:}
Not all candidate changes correspond to valid SM circuits, and candidate changes creating invalid SM circuits are excluded.
We verify SM circuit validity by checking \circlenumber{1} stabilizer commutation is preserved and \circlenumber{2} all CNOTs are schedulable.
For \circlenumber{2}, not schedulable CNOTs arise from cycles in the graph structure shown in Figure~\ref{fig:prop_graph} creating a circular dependence.

\textbf{Ambiguity Removal:}
Candidate changes originate from circuit-level errors, $E'$, involved in ambiguous errors found according to Sections~\ref{sec:ambig_finding} and ~\ref{sec:max_sat}.
We can therefore check if a candidate changes resolves the ambiguous errors.
This involves checking $L' \in \text{rowsp}(H')$ for updated circuit-level matrices $H', L'$ corresponding to the original ambiguous syndrome bits now in the candidate change's SM circuit.

We additionally need to check the updated circuit-level errors are not still a logical error.
Specifically, that $H'E' \neq 0$ or $L'E' = 0$. 

\subsection{Applying Verified SM Circuit Changes}
After pruning candidate SM circuit changes, we will have a set of verified changes that resolve the ambiguity found in Section~\ref{sec:ambig_finding}.
We attempt to apply all verified changes to the SM circuit, however, there may be instances where multiple changes are incompatible.
For example, if we find multiple verified circuit changes for a single ambiguous subgraph then applying one of them invalidates the ambiguity analysis for the others as the ambiguity in the subgraph has been removed.
We therefore only apply the change corresponding to the shortest depth SM circuit in such cases.
Circuit depth is a natural secondary optimization target here, as after Section~\ref{sec:prune} all changes resolve ambiguity and circuit depth is still relevant for minimizing decoherence errors.


\section{Evaluation}\label{sec:eval}
\begin{table}[t]
    \centering
    \caption{Selected QEC Codes for Evaluation}
    \label{tab:eval_code}
    \begin{tabular}{c|c}
                      \hline
     \textbf{Code Construction} & \textbf{Code Parameters} \\
                       \hline
                       & [[9,1,3]] \\
               Surface & [[25, 1, 5]]\\
                 Codes & [[49, 1, 7]]\\
                       & [[81, 1, 9]]\\
                       \hline
              LP Codes & [[39, 3, 3]]\\
                       \hline
                       & [[54, 11, 4]]\\
             RQT Codes & [[108, 18, 4]]\\
                       & [[60, 2, 6]]
    \end{tabular}
\end{table}

We evaluate PropHunt on a suite of QEC codes shown in Table~\ref{tab:eval_code}. 
Lifted Product (LP) and Random Quantum Tanner (RQT) codes are constructed using the qLDPC library~\cite{perlin2023qldpc}.
The $[[39, 3, 3]]$ LP code is constructed using the cyclic group $C_3$ and the protograph in Equation 8 of~\cite{roffe2023bias}.
The stabilizer matrices are a mix of weight 4, 5, and 6 stabilizers.
The $[[60, 2, 6]]$ RQT code is constructed from a length-2 repetition code and the cyclic group $C_{15}$.
The stabilizer matrices contain weight 4 stabilizers.
Finally the $[[54, 11, 4]]$ and $[[108, 18, 4]]$ RQT codes are both created using length-3 repetition codes and the Dihedral group $D_6$~\cite{radebold2025explicit}.
The stabilizer matrices contain weight 6 stabilizers.


\begin{figure*}
    \centering
    \includegraphics[width=\linewidth]{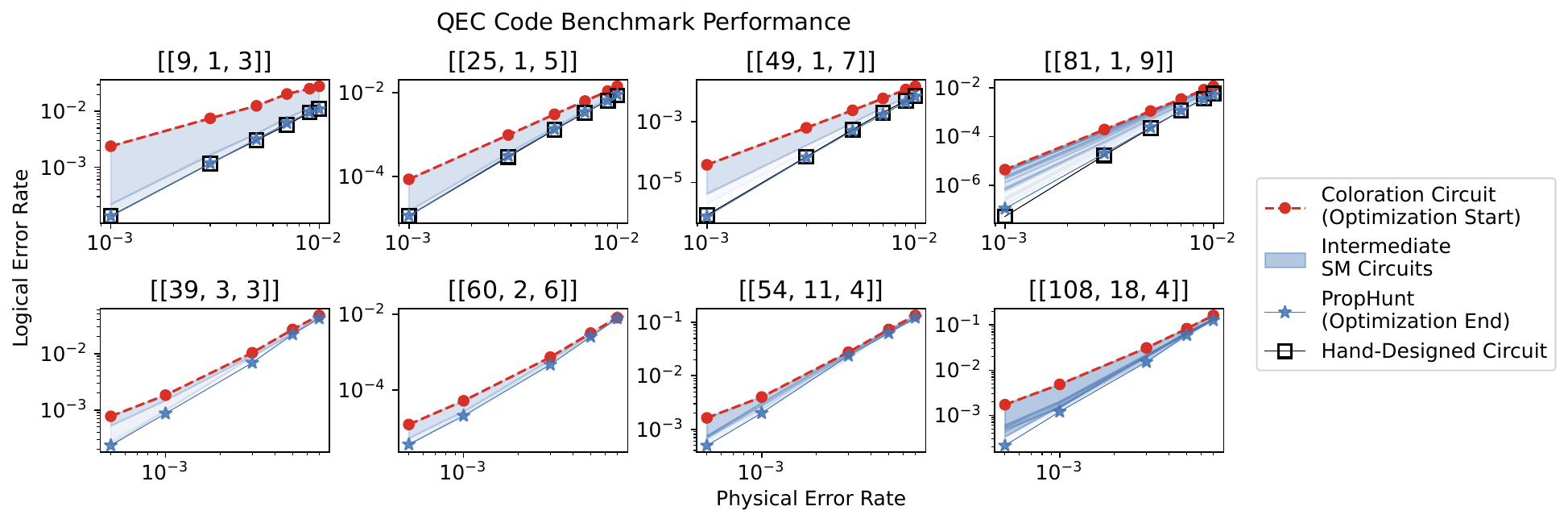}
    \caption{PropHunt's performance on benchmark QEC codes. Logical error rates include both logical $X$ and $Z$ error rates. For the surface codes, performance of a popular, hand-designed circuit is included for comparison.}
    \label{fig:qec_benchmarks}
\end{figure*}

\subsection{Optimizing Benchmark Codes}\label{sec:benchmark}

We \revision{first} discuss PropHunt's ability to optimize the benchmark QEC codes in Table~\ref{tab:eval_code}.

\revision{
We select the coloration circuit described in Algorithm 1 of~\cite{Tremblay2022} as our baseline. This is a generally applicable SM circuit technique for CSS codes and is currently employed by popular open-source libraries for circuit-level simulations~\cite{perlin2023qldpc, Roffe_LDPC_Python_tools_2022}.
}

All experiments start from the baseline circuit and ran for a maximum of 25 iterations with 500 samples per iteration.
Additionally, ambiguous subgraph finding was parallelized over 48 Intel Xeon Silver 4116 CPU cores.
Logical error rate simulations of the resulting SM circuits use a standard circuit-level model of $d$ rounds of the SM circuit.
\revision{
Single qubit operations have $\{X,Y,Z\}$ applied with probability $\frac{p}{3}$ after each gate and two qubit operations have $\{I,X,Y,Z\}\otimes\{I,X,Y,Z\}\setminus II$ applied with probability $\frac{p}{15}$ after each gate.
}
Simulations are done using Stim~\cite{gidney2021stim}.
Decoding for surface codes was performed using PyMatching~\cite{Higgott2025sparseblossom}, and decoding for LP and RQT codes was performed using BP-LSD~\cite{hillmann2024localized}.
All logical error rates include both logical $X$ and $Z$ error rates.

The results are shown in Figure~\ref{fig:qec_benchmarks}. 
We see that for all codes, PropHunt is able to improve logical error rate compared to the coloration circuit.
For the surface codes, we see PropHunt produces SM circuits that match the performance of well-known, hand-designed circuits, indicating PropHunt's ability to successfully automate SM circuit optimization.
For the LP and RQT codes where performant SM circuits are unknown, PropHunt successfully identifies and resolves problematic CNOT ordering and scheduling that causes reduced performance.
Compared to the coloration circuit, this improves logical error rates by 2.5x-4x at a physical error rate of $0.1\%$.

\begin{figure}
    \centering
    \includegraphics[width=\linewidth]{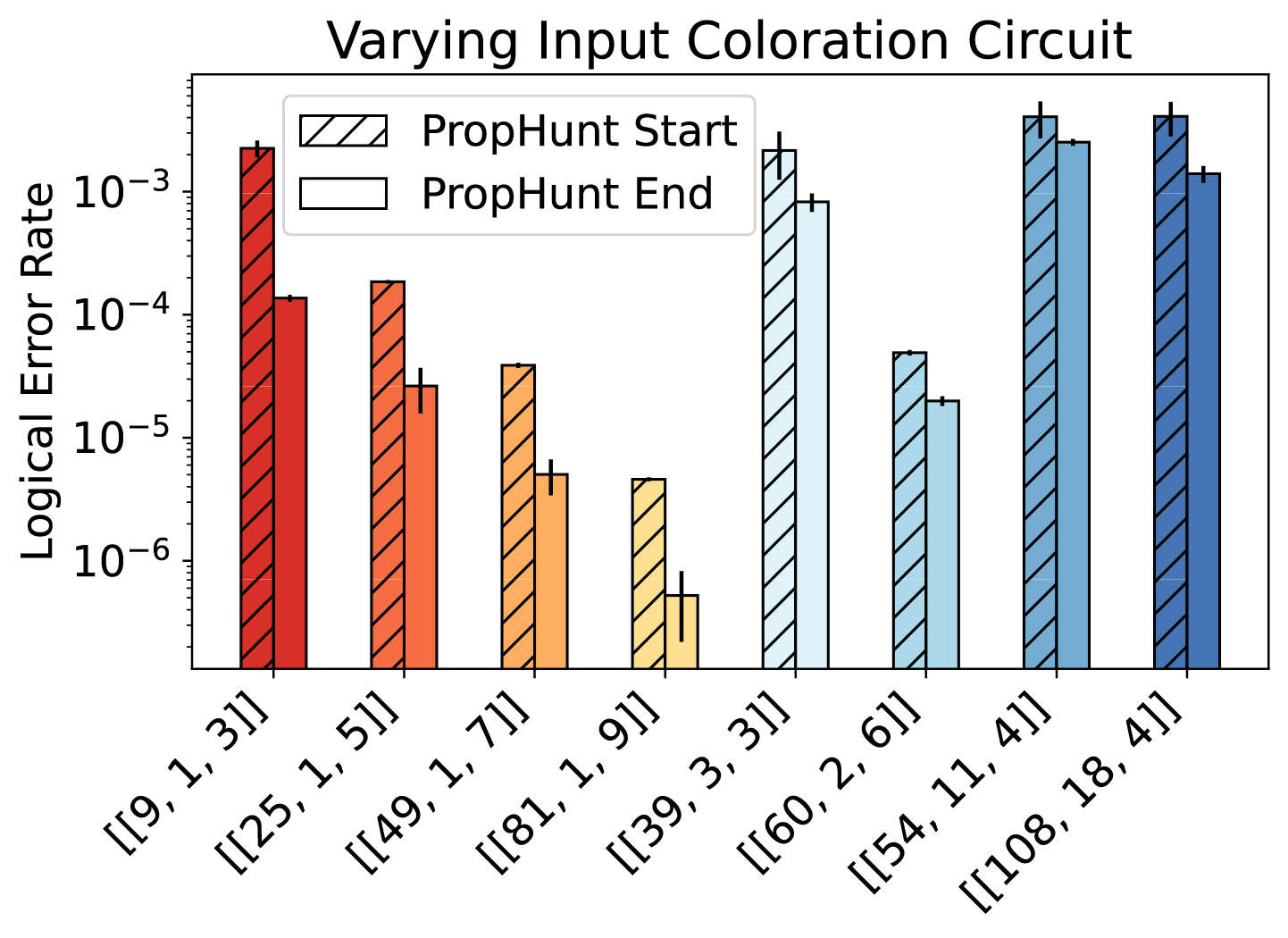}
    \caption{
    \revision{Evaluating PropHunt given three different, random coloration circuits as input. Data is for a physical error rate of $0.1\%$. Although variation exists in both starting and ending performance, indicated by vertical bars, PropHunt is able to consistently improve performance of the input coloration circuit.
    }}
    \label{fig:vary_input}
\end{figure}

\revision{
In Figure~\ref{fig:vary_input} we also evaluate PropHunt on three different, random coloration circuits. We observe similar performance as in Figure~\ref{fig:qec_benchmarks}, indicating robustness in PropHunt's ability to improve SM circuit performance for the benchmark codes we evaluate.
}

\subsection{Finding Min-Weight Ambiguous Errors}

\begin{table*}[t]
    \centering
    \begin{tabular}{c|c|c|c|c|c|c}
    \textbf{Formulation} & \textbf{Code} & \textbf{SM Circuit Distance} & \textbf{Variables} & \textbf{Hard Clauses} & \textbf{Soft Clauses} & \textbf{Wall Clock Time} \\
    \hline
                    & $$[[39,3,3]]$$ & $d_{\text{eff}}=2$ &             27958  &               128709  &                7884   &                101.36 s   \\
    Global          & $$[[49,1,7]]$$ & $d_{\text{eff}}=4$ &             45050  &               182533  &                16968  &                1 hr 55 min    \\
                    & $$[[60,2,6]]$$ & $d_{\text{eff}}=4$ &             60815  &               266381  &                20880   &               *   \\
    \hline \hline
                    & $$[[39,3,3]]$$ & $d_{\text{eff}}=2$ &             257    &               900     &                123    &                1.18 s   \\
    Subgraph        & $$[[49,1,7]]$$ & $d_{\text{eff}}=4$ &             340    &               1196    &                164    &                1.28 s    \\
                    & $$[[60,2,6]]$$ & $d_{\text{eff}}=4$ &             246    &               941    &                 110   &                 1.39 s   
    \end{tabular}
    \caption{Comparing example MaxSAT model size between global solving and ambiguous subgraph solving. * Solver timed out. }
    \label{tab:model_size}
\end{table*}

We \revision{now} discuss the performance of PropHunt's solution for finding ambiguous errors in an SM circuit. 
Since found ambiguous errors drive schedule changes in PropHunt, the performance of this stage is an optimization bottleneck.

Using just a code's stabilizer matrices, sampling is commonly used to estimate $d$~\cite{pryadko2023qdistrnd}.
Extending this to the circuit-level model, however, is not feasible due to the large check matrices involved as described in Section~\ref{sec:circuit_level}.
This also makes globally solving for min-weight logical errors in the circuit-level model intractable.
In Table~\ref{tab:model_size} we show MaxSAT model sizes based on the formulation in Section~\ref{sec:max_sat}.
Our ambiguous subgraph approach results in models much more tractable than a global approach, allowing for min-weight logical errors to be found efficiently.
Additionally, ambiguous subgraphs are sampled in parallel, allowing for multiple parts of a SM circuit to be optimized simultaneously.

\begin{figure}
    \centering
    \includegraphics[width=\linewidth]{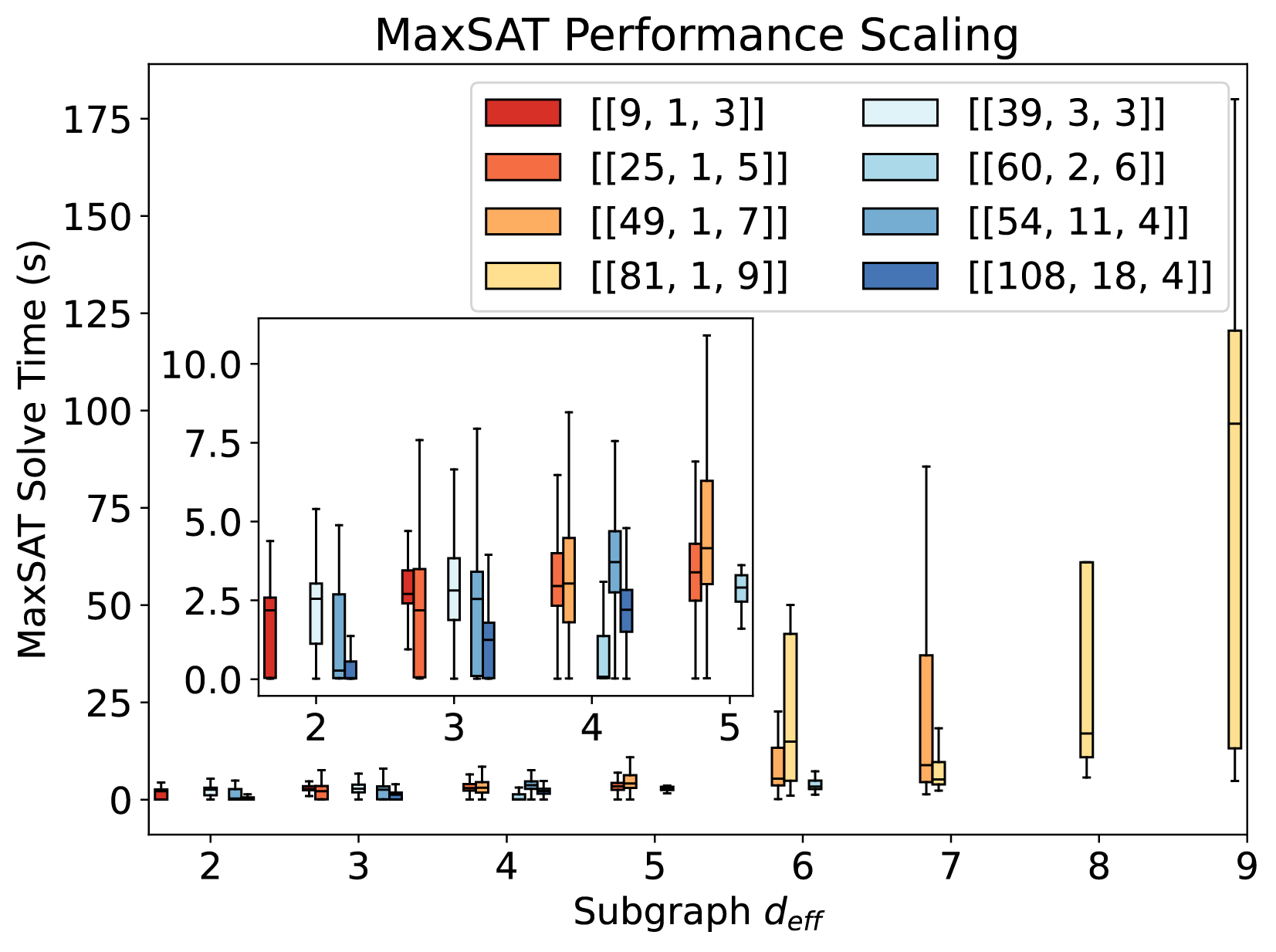}
    \caption{
    \revision{Scaling analysis of our ambiguous subgraph MaxSAT formulation. Data is collected from 1000 ambiguous subgraph samples during each iteration of PropHunt optimization. Model size and solve times both scale with $d_{\text{eff}}$ which increases during optimization, eventually saturating at $d_{\text{eff}} = d$ for the benchmark codes we consider. }}
    \label{fig:maxsat_scaling}
\end{figure}

\revision{
\subsubsection{Ambiguous Subgraph MaxSAT Scaling}
In Figure~\ref{fig:maxsat_scaling} we plot solve times for our ambiguous subgraph formulation during PropHunt optimization. Subgraph size scales with $d_{\text{eff}}$ and so we expect solve times to increase for larger $d_{\text{eff}}$. Interestingly, we observe high variability in solve times for certain codes. For example, at $d_{\text{eff}}=6$ the $d=7$ and $d=9$ surface codes have a notably larger range in solve times compared to the $d=6$ RQT code. This variability also increases at higher $d_{\text{eff}}$. For example, this effect is much less noticeable for the $d=3$ and $d=5$ surface codes, indicating solve time variability will likely increase with higher distance codes in the future.
}

\subsection{Idle Error Sensitivity Study}

\begin{figure}
    \centering
    \includegraphics[width=\linewidth]{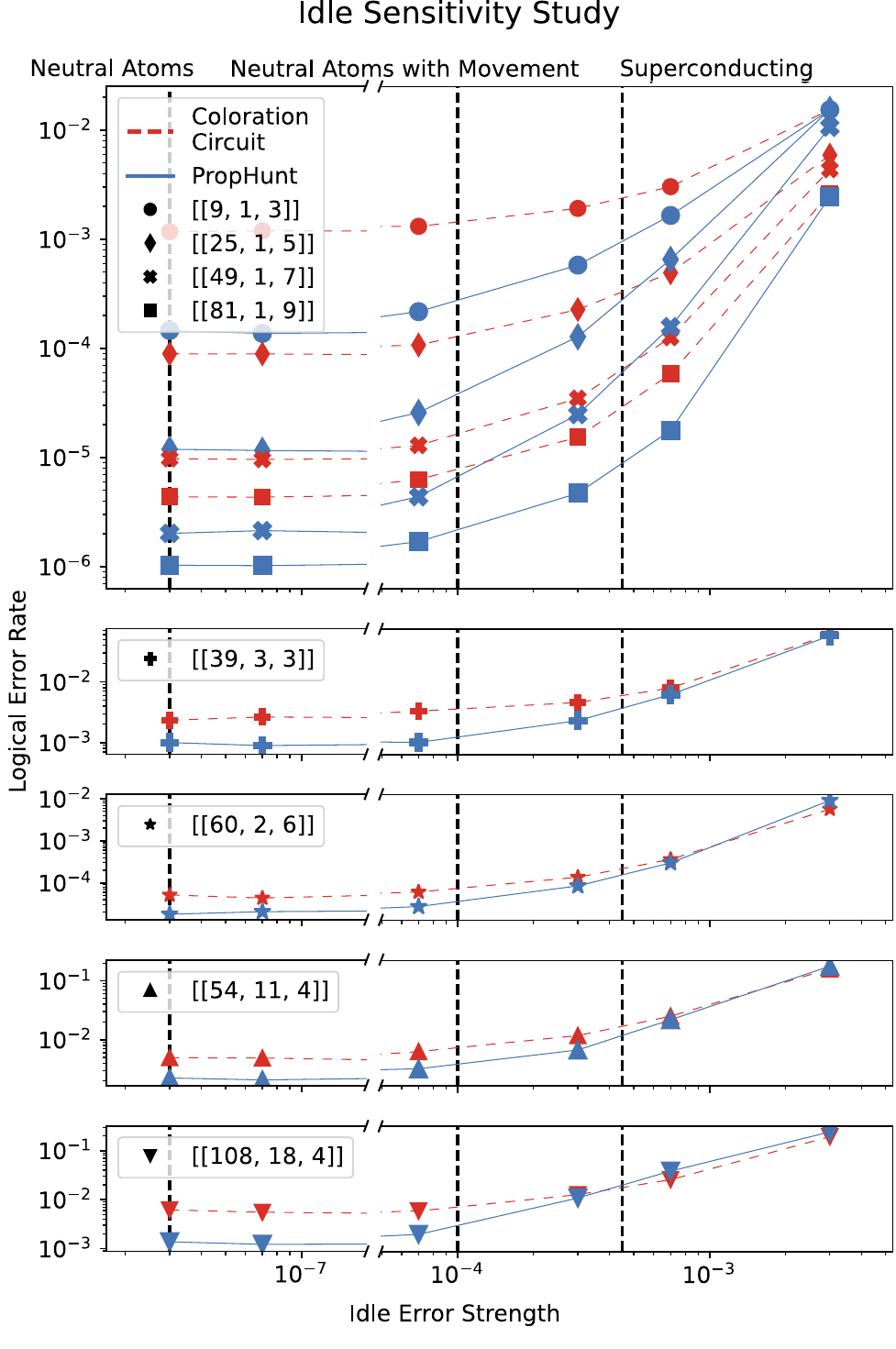}
    \caption{Idle error sensitivity study for benchmark QEC codes at a gate error rate of $0.1\%$. Idle error strength is the ratio of two-qubit gate time to qubit coherence time. For neutral atoms with movement, we assume movement is needed between each two-qubit gate layer.}
    \label{fig:idle_study}
\end{figure}

PropHunt is different than prior work in that it focuses on modifying error propagation in SM circuits to reduce ambiguity, and circuit depth is not its primary optimization target.
We believe this is well-motivated, as for many hardware platforms two-qubit gate layers are not bottlenecks.
For example, in neutral atoms, two-qubit gate times are $\sim 300$ ns, compared to measurement times of $\sim 1$ ms~\cite{radnaev2024universal, bluvstein2024logical}.
Small increases in circuit depth can therefore be preferable if they improve the resulting logical error rate.

To understand this trade-off, we study the sensitivity of the SM circuits evaluated in Section~\ref{sec:benchmark} to idle errors between gate layers.
We add idle errors to our simulations using a popular Pauli twirling approximation~\cite{tomita2014low}, and use a fixed gate error rate of $0.1\%$.
We additionally include three points of interest corresponding to neutral atom systems, superconducting systems, and movement-based neutral atom systems based off of experimental results~\cite{radnaev2024universal, google2025quantum, bluvstein2024logical, ma2022universal}.
For movement-based neutral atoms, we assume each gate layer requires $500 \mu s$ of movement, however, the exact amount of movement varies by code, with surface codes requiring less movement~\cite{bluvstein2022quantum}.
Movement costs could also be reduced using automated tools~\cite{lin2025reuse, stade2024abstract, tan2025compilation, wang2024atomique}.

The results are shown in Figure~\ref{fig:idle_study}.
Idle error strength is defined as the ratio: $t_g / T$ where $t_g$ is the idle time for each gate layer and $T$ is the qubit coherence time.
In general, we can see that for a large range of relevant idle error strengths, small increases in circuit depth from PropHunt are outweighed by the improvements in logical error rate.
We leave further evaluation of codes for specific hardware to future work, and hope the data in Figure~\ref{fig:idle_study} is informative to a broad range of experimental setups.

\section{Fine-Grained Noise Scaling for QEC–ZNE}

QEC is essential for suppressing physical noise but is costly and, on near-term hardware, unlikely to achieve demanding fidelity targets on its own.
\revision{A unique feature of the results in Section~\ref{sec:eval} is the intermediate logical error rates produced by PropHunt during optimization. We note that more intermediate logical error rates can easily be achieved by decreasing the samples per iteration and increasing the number of iterations, slowing down optimization convergence.
We find this is a natural fit to Zero-Noise Extrapolation (ZNE)~\cite{temme2016error, li2017efficient}, an error mitigation technique which intentionally increases error rates to then fit expectation values to an estimated zero-noise limit. Given this insight, we propose a ZNE-based application of PropHunt, termed \textit{Hook-ZNE}.  
Since ZNE is a complementary, hardware-agnostic mitigation technique we find this promising for near-term applications of PropHunt.
}



\subsection{Challenges in Current QEC–ZNE Integration}
\revision{
With QEC The logical error rate scales as
\small
\begin{equation*}
  P_L(d) \;\approx\; \left(\frac{P}{P_{\text{th}}}\right)^{\lceil\frac{d}{2}\rceil}
  \;=\; \Lambda^{-\lceil\frac{d}{2}\rceil},
  \qquad
  \Lambda \equiv \frac{P_{\text{th}}}{P},
\end{equation*}
\normalsize
where \(d\) is the code distance, \(P\) is the physical error rate, and \(P_{\text{th}}\) is the threshold error rate. When the hardware operates below threshold (\(P < P_{\text{th}}\)), we have \(\Lambda > 1\), so increasing the code distance \(d\) suppresses the logical error rate \(P_L(d)\) exponentially. Utilizing this relationship among \(P_L(d),~\Lambda,\) and \(~d \), prior works combine QEC with ZNE, yet all face challenges in effectively reducing errors.} 

\textbf{Logical Circuit Folding} increases noise by appending inverse–forward logical blocks, effectively stretching the circuit depth. While this provides a tunable way to amplify errors, it also lengthens the execution time, which is especially costly on error-corrected hardware where operations are already slow and resource-intensive.

\textbf{Distance-Scaling ZNE (DS-ZNE)} scales the code distance to vary logical error rates~\cite{wahl2023zero}. If an application can run at distance $d$, it is also executed at $d-2, d-4, \ldots, d-2k$. Lowering $d$ increases noise, enabling extrapolation back to zero-noise. However, two limitations arise: \circlenumber{1} the scaling factors are coarse, since logical error rates fall exponentially with linear increases in $d$, and distance values are restricted to odd integers; \circlenumber{2} on near-term devices with small maximum distances, the variance of observables grows quickly at low $d$, and only a handful of $d$ values are available, making fitting unreliable. \circlenumber{3} Scaling Code distance requires complete recompilation, including distance-specific mapping, routing, and Clifford + T synthesis optimizations. 
In theory, DS-ZNE enables higher throughput by lowering code distance and freeing qubits to run parallel copies of the computation. In practice, however, most qubits are consumed by magic state distillation, which already operates at low distance utilizing post-selection. As a result, reducing the distance of algorithmic qubits does little to free resources, making the parallelism advantage largely negligible.

    \textbf{Physical-Level Noise Amplification} modifies physical gate durations or idle times at fixed $d$ to amplify physical noise within SM circuits~\cite{zhang2025demonstrating}. Unfortunately, this would require fine-grained control over physical gates used for SM. Moreover, amplifying noise without any bias at the physical level is challenging and can lead to nonlinear noise scaling and distortion of error structure, especially when the physical error rate is close to the threshold. Another challenge in scaling physical errors is increased decoding pressure due to higher physical error rates, which can make decoding more challenging and lead to unintended decoder slowdown.


\begin{figure}
    \centering
    \includegraphics[width=\linewidth]{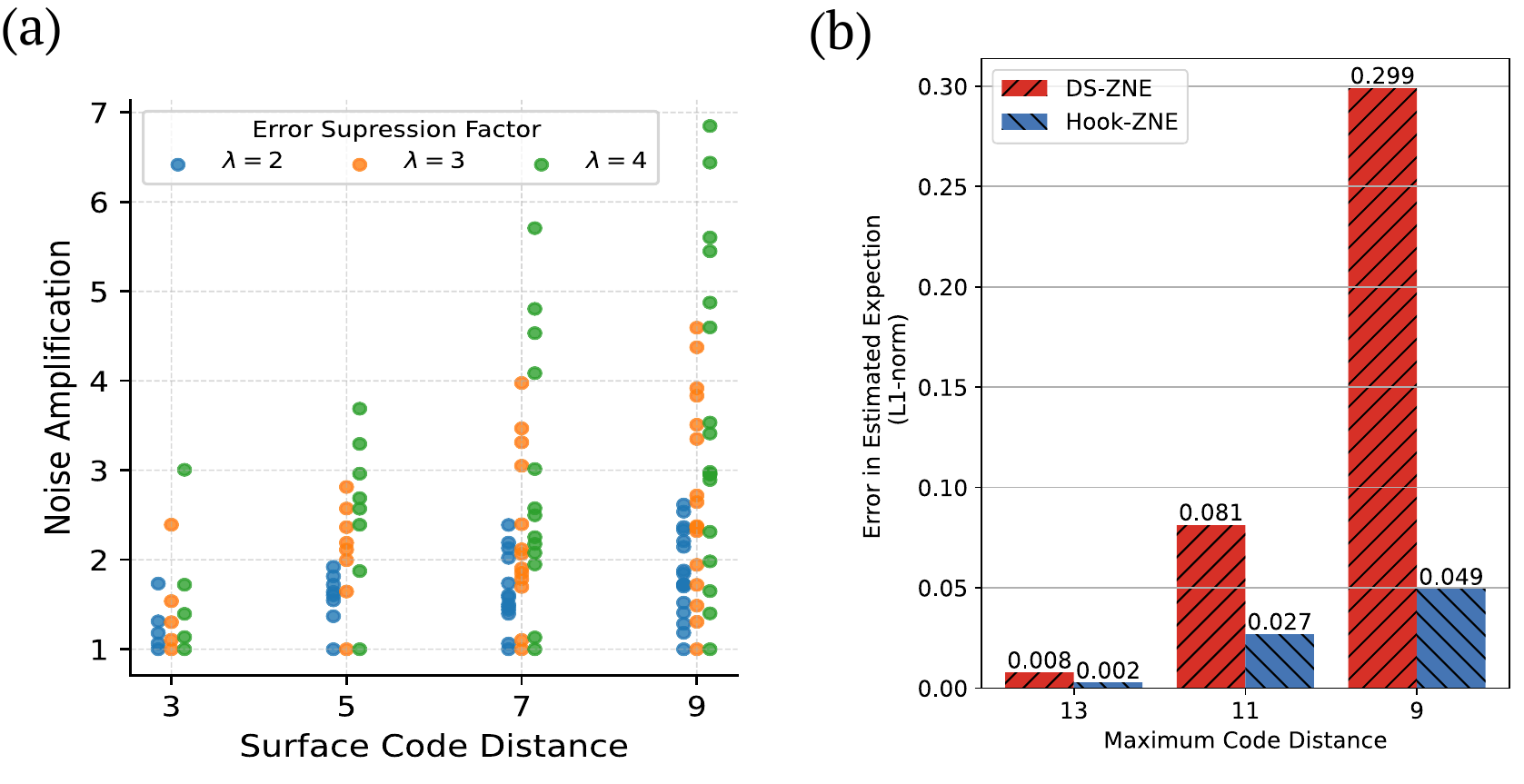}
    \caption{(a) Noise Amplification with Hook-ZNE (b) Bias in expectation values for DS-ZNE vs. Hook-ZNE.}
    \label{fig:zne_noiseamp_all}
\end{figure}

\subsection{Hook-ZNE Design and Evaluation}

\revision{Hook-ZNE amplifies logical noise smoothly at fixed code distance by leveraging suboptimal SM circuits discovered during PropHunt's optimization.}
This provides fine-grained, low-overhead control over the logical error rate without increasing logical circuit depth or qubit count, enabling bias–variance trade-offs that are better suited for near-term, resource-constrained fault-tolerant quantum computing. 
Figure~\ref{fig:zne_noiseamp_all}a illustrates the range of noise amplification achievable for a surface code at fixed code distance under different suppression factors, $\Lambda$. 
For example, in Google’s surface code experiments, the reported suppression factor was $\Lambda = 2.14$~\cite{google2025quantum}.

We further evaluate \textbf{Hook-ZNE} against \textbf{DS-ZNE} by comparing the \emph{bias}, defined as the \(L_1\) distance between the estimated and ideal expectation value of the target observable. Following the DS-ZNE setup with suppression factor \(\Lambda = 2\), we test three distance ranges. For DS-ZNE, we use distances with odd integers. For Hook-ZNE, we round to finely spaced values of $d$ \revision{which can be understood as fits to $ P_L(d) = \Lambda^{-\frac{d+1}{2}}$}:

\begingroup\small
\setlength{\jot}{1pt}
\[
\begin{aligned}
\text{DS-ZNE: } & d\in\{13,11,9,7;\;11,9,7,5;\;9,7,5,3\}\\
\text{Hook-ZNE: } & d\in\{13,12.5,12,11.5;\;11,10.5,10,9.5;\;9,8.5,8,7.5\}
\end{aligned}
\]
\endgroup

\revision{We use mitiq~\cite{larose2022mitiq} to generate randomized benchmarking circuits with a two-qubit gate circuit depth of 50. Uniform gate noise is added for each gate with magnitude defined by $ P_L(d) = \Lambda^{-\frac{d+1}{2}}$. To compare DS-ZNE versus Hook-ZNE, we use a total budget of 20,000 shots. The results are shown in Figure~\ref{fig:zne_noiseamp_all}b. We can see Hook-ZNE consistently produces estimations closer to the true expectation value across all configurations.}
We find the key advantage of Hook-ZNE is its \emph{finer noise scaling} at essentially fixed \(d\), which minimizes exposure to very low distances where noise amplification and estimator variance grow rapidly. This leads to more stable fits under the same total shot budget. 



\section{Related Work}
\revision{
\subheading{Alternative approaches for finding SM circuits}

An alternative approach for finding performant CNOT schedules in SM circuits is to \circlenumber{1} identify repetitive code structures to create a smaller, parameterized circuit and \circlenumber{2} perform a brute-force search over the parameterized circuit.
This strategy has been successfully employed for bivariate bicycle codes~\cite{bravyi2024high} and color codes~\cite{lee2025color}.

Another approach is to design SM circuits by hand, which has been successfully employed for surface codes~\cite{tomita2014low} and generalized bicycle codes~\cite{lin2025single}.

\subheading{Flag fault-tolerant SM circuits}

There is related work that uses extra flag qubits to detect hook errors and create SM circuits where $d_{\text{eff}}=d$~\cite{Chamberland2018, Chao2018, chao2020flag, Chamberland2020}. 
Flag-Proxy Networks~\cite{vittal2024flag} is an automated approach using flag qubits and proxy qubits for creating distance-preserving SM circuits that also have low connectivity requirements.
In ~\cite{Shutty2022} the authors used SMT solvers and flag fault-tolerance to create fault-tolerant circuits for preparing $|H\rangle$-type magic states in the color code.


In contrast to these related works, PropHunt does not use extra ancilla qubits to detect hook errors. However, future work could explore augmenting the circuits output by PropHunt with flag fault-tolerance.

\subheading{Related SM circuit problems}
}

QECC-synth~\cite{Yin2025} addressed SM circuit compilation for QEC codes that are natively incompatible with the underlying hardware architecture. This work extended and improved other prior work that used ancilla qubits for enabling QEC codes on incompatible hardware architectures~\cite{Wu2022}.
Gehér et al. also introduced a tangling schedule that reduced hardware connectivity requirements for QEC codes~\cite{Gehr2024}.

\revision{
In ~\cite{Berthusen2025} the authors proposed an adaptive SM circuit strategy for concatenated codes where the outer codes are measured less frequently, dependent on logical soft information from the inner codes.

Automated approaches also exist for creating surface code SM circuits in the presence of defective components using high-weight superstabilizers comprised of low-weight gauge operators~\cite{siegel2023adaptive, lin2024codesign, yin2024surf, leroux2024snakes, debroy2025luci, higgott2025handling}.

In theory, PropHunt is complementary to these related SM circuit problems and solutions, but thoroughly exploring the connection is a step we leave for future work.




\subheading{Alternative approaches to syndrome measurement}

Other methods of measuring syndromes exist including Shor-style~\cite{shor1996fault}, Steane-style~\cite{steane1997active}, and Knill-style~\cite{knill2005quantum}. In contrast to the SM circuit method, these methods cost extra logical ancilla qubits. These logical ancilla qubits must be prepared using a state preparation circuit which is addressed in related work~\cite{peham2025automated, zen2025quantum}.

}





\section{Conclusion}
In this work we addressed SM circuit optimization for \revision{CSS} QEC codes.
We point out imperfect performance predictors used in prior work, and introduce a novel approach based on minimizing ambiguity in SM circuits.
We present a tool, PropHunt, for optimizing SM circuits and find PropHunt produces SM circuits that recover the performance of hand-designed circuits.
Additionally, circuits for LP and RQT codes produced by PropHunt have 2.5x-4x lower logical error rates than existing circuits.
Finally, we show intermediate SM circuits created by PropHunt can be used for error mitigation with ZNE, which we call Hook-ZNE.
Our evaluations show Hook-ZNE reduces error by 3x-6x compared to DS-ZNE.
\begin{acks}
    We would like to thank the anonymous reviewers for their helpful feedback and suggestions.

    This work is funded in part by the STAQ project under award NSF Phy-232580; in part by the US Department of Energy Office of Advanced Scientific Computing Research, Accelerated Research for Quantum Computing Program; in part by the NSF Quantum Leap Challenge Institute for Hybrid Quantum Architectures and Networks (NSF Award 2016136); in part by the NSF National Quantum Virtual Laboratory program; in part based upon work supported by the U.S. Department of Energy, Office of Science, National Quantum Information Science Research Centers; in part by the Army Research Office under Grant Number W911NF-23-1-0077; in part by the U.S. Department of Energy, Office of Science, National Quantum Information Science Research Centers, Co-design Center for Quantum Advantage (C2QA) under contract number DE-SC0012704; and in part by the U.S. National Science Foundation (NSF) under Award Abstract number 2212232.
    
    The views and conclusions contained in this document are those of the authors and should not be interpreted as representing the official policies, either expressed or implied, of the U.S. Government. The U.S. Government is authorized to reproduce and distribute reprints for Government purposes notwithstanding any copyright notation herein.
    FTC is the Chief Scientist for Quantum Software at Infleqtion.
\end{acks}

\revision{
\section*{Data Availability}
The code and data for this work is open-source and available at \href{https://github.com/jviszlai/PropHunt}{github.com/jviszlai/PropHunt}.

}
\bibliographystyle{ACM-Reference-Format}
\balance
\bibliography{ref}

%
%
%
%
%


\appendix
\section{Artifact Appendix}

\subsection{Abstract}

The artifact includes the source code, data, and scripts necessary to reproduce the results in this paper. Additionally, it contains our introduced tool, PropHunt, for optimizing syndrome measurement circuits of CSS codes.

\subsection{Artifact check-list (meta-information)}

All details, including software and hardware requirements, are summarized below:

{\small
\begin{itemize}
  \item {\bf License}: MIT License
  \item {\bf Software requirements:} Docker, Python
  \item {\bf Hardware:} Multi-core CPU
  \item {\bf How much disk space required (approximately)?:} 2GB
  \item {\bf How much time is needed to prepare workflow (approximately)?:} 5-10 minutes to build Docker image
  \item {\bf How much time is needed to complete experiments (approximately)?:} 8-12 hours depending on number of CPU cores
  \item {\bf Publicly available?:} Yes
  \item {\bf Archived (provide DOI)?:} 10.5281/zenodo.17945386
\end{itemize}
}

\subsection{Description}

\subsubsection{How to access}

The most up-to-date version of the code can be accessed by cloning the repository from GitHub:
\begin{lstlisting}[style=terminalstyle]
git clone https://github.com/jviszlai/PropHunt.git
\end{lstlisting}

Alternatively the archived artifact version can be accessed by downloading and unzipping the Zenodo archive.

\subsection{Installation}

To simplify the building and integration of the Loandra MaxSAT solver into PropHunt, we've provided a Docker image for data generation. We've also created a Makefile to simplify evaluation.

The Docker image can be created most easily by running:
\begin{lstlisting}[style=terminalstyle]
make docker
\end{lstlisting}
from the main directory.

Plotting uses Python. \texttt{numpy} and \texttt{matplotlib} are required.

\subsection{Experiment workflow}

Scripts for running experiments are provided in the \texttt{scripts} directory.
Running all experiments using the Docker image can be done by:
\begin{lstlisting}[style=terminalstyle]
make data
\end{lstlisting}
from the main directory. Produced data is mounted to the local \texttt{scripts/data} directory to be used outside the Docker container. 

\texttt{scripts/plot\_data.py} will generate figures for the manuscript in the \texttt{scripts/figures} directory.
This can also be done by:
\begin{lstlisting}[style=terminalstyle]
make plots
\end{lstlisting}
from the main directory.

\subsection{Evaluation and expected results}

Simulations should generate results for:
\begin{itemize}
    \item Motivational data on distance and $d_{\text{eff}}$ as performance predictors
    \item Hand-optimized surface code SM circuit performance
    \item MaxSAT solver timing results for codes in Table~\ref{tab:model_size}
    \item PropHunt results for benchmark codes compared to coloration circuit
    \item Sensitivity study to idle errors
    \item Comparison between DS-ZNE and Hook-ZNE
\end{itemize}
Plotting should generate plots or subplots for figures 1, 12, 15, and 16.

\subsection{Experiment customization}
\texttt{NUM\_CORES} in \texttt{run\_experiments.sh} should be changed to the number of cores on the system.

Additionally, the number of random samples per iteration, the number of iterations, and the number of cores used in PropHunt can be customized per benchmark in \texttt{run\_experiments.sh}. For example:
\begin{lstlisting}[style=terminalstyle]
python prophunt_experiment.py surface 3 100 5 48
\end{lstlisting}
specifies the $d=3$ surface code benchmark, with 100 samples per iteration, for 5 iterations, run on 48 cores.






\end{document}